\documentclass[twocolumn]{aastex7}

\usepackage{enumitem}
\usepackage{CJK}
\usepackage{pifont}
\usepackage{float}

\newcommand\be{\begin{eqnarray}}
\newcommand\ee{\end{eqnarray}}
\newcommand{\Rmnum}[1]{\uppercase\expandafter{\romannumeral #1}}

\shorttitle{Hot gas origins in MW-like galaxies}
\shortauthors{Zhang et al.}

\begin{document}
\begin{CJK*}{UTF8}{gbsn}

\title{Tracing the Origins of Hot Halo Gas in Milky Way-Type Galaxies with SMUGGLE}

\correspondingauthor{Xiaoxia Zhang; Hui Li; Taotao Fang}
\email{zhangxx@xmu.edu.cn; hliastro@tsinghua.edu.cn; fangt@xmu.edu.cn}

\author[0000-0002-8552-2558]{Zhijie Zhang (张志杰)}
\affiliation{Department of Astronomy, Xiamen University, Xiamen, Fujian 361005, People's Republic of China}
\email{}
\author[0000-0003-4832-9422]{Xiaoxia Zhang (张小霞)}
\affiliation{Department of Astronomy, Xiamen University, Xiamen, Fujian 361005, People's Republic of China}
\email[]{zhangxx@xmu.edu.cn}  
\author[0000-0002-1253-2763]{Hui Li (李辉)}
\affiliation{Department of Astronomy, Tsinghua University, Beijing 100084, People's Republic of China}
\affiliation{Department of Astronomy, Columbia University, Manhattan, New York 10027, USA}
\email[]{hliastro@tsinghua.edu.cn}  
\author[0000-0002-2853-3808]{Taotao Fang (方陶陶)}
\affiliation{Department of Astronomy, Xiamen University, Xiamen, Fujian 361005, People's Republic of China}
\email[]{fangt@xmu.edu.cn}  
\author[0000-0002-2243-2790]{Yang Luo (罗阳)}
\affiliation{Department of Astronomy, Yunnan University, Kunming, Yunnan 650000, People's Republic of China}
\email{}
\author[0000-0003-3816-7028]{Federico Marinacci}
\affiliation{Department of Physics and Astronomy "Augusto Righi", University of Bologna, via Gobetti 93/2, I-40129 Bologna, Italy}
\affiliation{INAF, Astrophysics and Space Science Observatory Bologna, Via P. Gobetti 93/3, I-40129 Bologna, Italy}
\email{}
\author[0000-0002-3790-720X]{Laura V. Sales}
\affiliation{Department of Physics and Astronomy, University of California, Riverside, 900 University Avenue, Riverside, CA 92521, USA}
\email{}
\author[0000-0002-5653-0786]{Paul Torrey}
\affiliation{Department of Astronomy, University of Virginia, 530 McCormick Road, Charlottesville, VA 22903, USA}
\email{}
\author[0000-0001-8593-7692]{Mark Vogelsberger}
\affiliation{Department of Physics and Kavli Institute for Astrophysics and Space Research, Massachusetts Institute of Technology, Cambridge, MA 02139, USA}
\email{}
\author[0000-0003-3230-3981]{Qingzheng Yu (余清正)}
\affiliation{Dipartimento di Fisica e Astronomia, Universit\`a degli Studi di Firenze, Via G. Sansone 1, 50019 Sesto Fiorentino, Firenze, Italy}
\email{}
\author[0000-0003-3564-6437]{Feng Yuan (袁峰)}
\affiliation{Center for Astronomy and Astrophysics and Department of Physics, Fudan University, Shanghai 200438, People's Republic of China}
\email{}
\begin{abstract}

Current galaxy formation models predict the existence of X-ray-emitting gaseous halos around Milky Way (MW)--type galaxies. To investigate properties of this coronal gas in MW-like galaxies, we analyze a suite of high-resolution simulations based on the SMUGGLE framework and compare the results with X-ray observations of both the MW and external galaxies. We find that for subgrid models incorporating any form of stellar feedback, e.g., early feedback (including stellar winds and radiation) and/or supernova (SN) explosions, the total 0.5--2 keV luminosity is consistent {\it within uncertainties} with X-ray observations of the MW and with scaling relations derived for external disk galaxies. However, all models exhibit an X-ray surface brightness profile that declines too steeply beyond $\sim5$\,kpc, underpredicting the extended emission seen in recent eROSITA stacking results. Across all subgrid prescriptions, the simulated surface brightness and emission measure fall below MW observations by at least 1--2 orders of magnitude, with the most severe discrepancy occurring in the no-feedback model. Our results suggest that (i) stellar feedback primarily shapes the innermost hot atmosphere (central $\sim5$\,kpc), with comparable contributions from early feedback and SNe to the resulting X-ray luminosity; (ii) additional mechanisms such as gravitational heating, active galactic nuclei feedback, and/or Compton effects of GeV cosmic ray are necessary to generate the extended, volume-filling hot gaseous halo of MW-mass galaxies; (iii) the origins of hot corona in MW-like galaxies are partially distinct from those of the warm ($\sim10^5$\,K) gas, by combining our previous finding that the SMUGGLE model successfully reproduces the kinematics and spatial distribution of MW \ion{O}{6} absorbers.

\end{abstract}


\section{Introduction}
The presence of extended X-ray coronae around massive galaxies has long been predicted by galaxy formation theory \citep[e.g.,][]{1956ApJ...124...20S, 1978MNRAS.183..341W, 1991ApJ...379...52W}. In the standard picture of galaxy formation,  infalling gas  from the intergalactic medium (IGM) experiences accretion shocks, heating it to the galaxy's virial temperature ($T\gtrsim10^6$\,K). 
For massive ($M_{\rm vir} \gtrsim 10^{12}\,M_\odot$) galaxies, the cooling time of the shock-heated gas is sufficiently long to establish a quasi-hydrostatic equilibrium, forming a volume-filling, X-ray-emitting gaseous halo \citep[e.g.,][]{2024ApJ...974..291O}. 
Lower-mass galaxies, due to their shallower potential wells, are generally expected to host fainter X-ray coronae. The properties of their hot gas are expected to be shaped by feedback from stellar winds, supernovae \citep[SNe; e.g.,][]{2004ApJ...606..829S, 2010MNRAS.407.1403C, 2012MNRAS.421...98D, 2018ApJ...867...73S}, and active galactic nuclei \citep[AGN; e.g.,][]{2020MNRAS.494..549T}.   
However, for Milky Way (MW) analogues ($M_{\rm vir}\sim10^{12}\,M_\odot$), which lie at the boundary between these mass regimes, there is no consensus on the dominant heating mechanism.

Observationally, the hot gaseous halo of our Galaxy has been detected both in emission as a component of the soft X-ray background \citep[e.g.,][]{2013ApJ...773...92H, 2024ApJS..271...62P}, and in absorption (e.g., \ion{O}{7} and \ion{O}{8}) in high-resolution X-ray spectra of background sources \citep[e.g.,][]{2003ApJ...586L..49F, 2006ApJ...644..174F, 2015ApJS..217...21F, 2018ApJS..235...28L}. 
These observations yield a characteristic temperature of $T \approx 2\times10^6$\,K, though this value remains model-dependent.
With high spatial resolution of Chandra, extra-planar X-ray emission has been detected for external, mostly edge-on disk galaxies, to minimize point-source contamination. Those observations targeted galactic central regions, revealing typical scale heights of 1--30\,kpc, with a median value of $\sim5$\,kpc \citep[e.g.,][]{2004ApJ...606..829S, 2006A&A...448...43T, 2013MNRAS.428.2085L}. 
Furthermore, tight correlations were revealed between X-ray luminosity ($L_{\rm X}$) and star formation rate (SFR), indicating hot gas origins associated with stellar activity \citep[e.g.,][]{2013MNRAS.435.3071L, 2016MNRAS.457.1385W}. 
However, the $L_{\rm X}$--SFR relation may alternatively reflect an underlying $L_{\rm X}$--$M_{\rm vir}$ correlation, as the growth rate of stellar mass is strongly coupled to the radiative cooling rate regulated by gravitational potential \citep[e.g.,][]{2013MNRAS.435.3071L}. Consequently, gravitational heating and stellar feedback remain observationally degenerate based solely on scaling relations.

The launch of eROSITA in 2019 has enabled systematic  measurements of X-ray emission in the circumgalactic medium (CGM) via stacking analysis of galaxy samples binned by  properties such as the stellar mass ($M_*$) and specific SFR \citep[sSFR;][]{2022A&A...666A.156C, 2022ApJ...936L..15C, 2024A&A...690A.267Z}. 
For MW-mass galaxies, this approach has revealed extended X-ray emission out to the virial radius ($\gtrsim 100$\,kpc), with tentatively lower X-ray luminosities compared to both star-forming galaxies and more massive systems \citep[e.g.,][]{2022A&A...666A.156C, 2025A&A...693A.197Z}.

 A number of simulation studies have explored the properties of galactic coronal gas, e.g., X-ray  luminosity, surface brightness, and scaling relations, using both cosmological hydrodynamical simulations \citep[e.g.,][]{2010MNRAS.407.1403C, 2015ApJ...804...72B, 2021MNRAS.502.2934K, 2022ApJ...936L..15C, 2024ApJ...974..291O} and isolated high-resolution simulations \citep[e.g.,][]{2020ApJ...898..148L, 2024MNRAS.531.2757J},  primarily adopting external viewing perspectives.  
State-of-the-art simulations incorporating diverse physical processes can broadly reproduce certain observed coronal properties. 
However, large-scale simulations (e.g., IllustrisTNG and EAGLE) often exhibit an X-ray dichotomy at $z\sim0$ driven by AGN feedback:
at a transitional stellar mass of $M_* \sim 10^{10.5-11}\,M_\odot$, star-forming galaxies show elevated soft X-ray luminosities compared to quiescent systems \citep[][]{2020MNRAS.494..549T, 2020ApJ...893L..24O}. 
In contrast, isolated simulations enable precise modeling of SN-driven outflows but typically neglect AGN feedback and gas accretion from the IGM. 
These model-dependent limitations lead to divergent conclusions on fundamental issues, notably whether stellar feedback alone can explain hot corona origins \citep[e.g.,][]{2015ApJ...800..102H, 2015ApJ...813L..27P}.  

Successful implementation of stellar feedback subgrid models in high-resolution zoom-in simulations offers critical insights into the origins of hot coronae in MW-type galaxies. 
The Stars and MUltiphase Gas in GaLaxiEs \citep[SMUGGLE;][]{2019MNRAS.489.4233M} is a physically motivated subgrid model for the ISM and stellar feedback that has reproduced key properties of stellar populations and cold ($T\lesssim10^4$\,K) interstellar gas \citep[][]{2020MNRAS.499.5862L, 2022MNRAS.514..265L, 2022MNRAS.517....1S, 2023MNRAS.520..461J, 2024MNRAS.529.4073L}. Notably, our previous work \citep{2024ApJ...962...15Z} established that the SMUGGLE model accurately simulates Galactic warm  ($T\sim10^5$\,K) gas traced by \ion{O}{6} absorption, capturing diagnostics such as scale heights and column density--line width relation.
Here, we extend this validation to hot coronae ($T\sim10^6$\,K) in MW-mass systems, 
examining simulated X-ray properties from both external and solar (inside-out) perspectives.
Given that the simulations model isolated disks with stellar feedback but exclude AGN activity and cosmological gas accretion, this work specifically tests whether stellar feedback alone can account for the observed X-ray signatures of halo gas. A comparison with observations allows us to assess whether additional heating mechanisms are necessary to generate extended, volume-filling hot coronae.

The paper is organized as follows. Section~\ref{sec:model} describes the SMUGGLE model and the simulation suite analyzed in this work. Section~\ref{sec:result} presents  hot gas properties, including X-ray luminosity, surface brightness, emission measure, and temperature. These results are further discussed in Section \ref{sec:discuss}, and in Section~\ref{summary} we summarize our conclusions.

\section{The SMUGGLE Model and Simulations}
\label{sec:model}

The high-resolution simulations we analyzed in this work were generated by a suite of runs performed   with the SMUGGLE framework \citep{2019MNRAS.489.4233M} as described in \citet{2020MNRAS.499.5862L}. SMUGGLE is a physically motivated subgrid model for the ISM and stellar feedback implemented within the moving-mesh code AREPO \citep{2010MNRAS.401..791S}. Below, we provide a brief overview of the model and simulation setup; detailed descriptions can be found in the original papers. 

The SMUGGLE model incorporates various physical processes, including gravity, hydrodynamics, gas cooling and heating, star formation, and stellar feedback.  Star particles are formed in cold, dense, and self-gravitating molecular gas reaching a density threshold of $n_{\rm th}=100\, \rm cm ^{-3}$. The local SFR for star-forming gas cells is governed by the star formation efficiency ($\epsilon_{\rm ff}$) per freefall time, expressed as $\dot{M}_*=\epsilon_{\rm ff} M_{\rm gas}/\tau_{\rm ff}$, where $M_{\rm gas}$ is the gas mass and $\tau_{\rm ff}$ is the freefall time of the gas cell.  

The model implements diverse channels of stellar feedback, categorized into two main types: (i) SN feedback, which injects substantial energy and momentum into the ISM;  and (ii) early feedback, encompassing radiative processes (e.g., photoionization and radiation pressure) from massive stars, as well as stellar winds originating from both young OB stars and older asymptotic giant branch (AGB) stars. Feedback energy and momentum are deposited into the 64 nearest gas cells, weighted by the solid angle subtended by each cell relative to the star particle.

The high-resolution simulations model an isolated galactic disk. This disk consists of an MW-sized galaxy with a total mass of $1.6\times10^{12}\,M_\odot$. The galaxy's mass components include a stellar bulge ($1.5\times10^{10} M_\odot$), a stellar disk ($4.7\times10^{10} M_\odot$), a gaseous disk ($9\times10^{9} M_\odot$), and a dark matter halo. The masses of these components are similar to those of the MW \citep[see][and references therein]{2016ARA&A..54..529B}. The gaseous disk density decreases exponentially with a scale length of $6$\,kpc.
The mass resolution of the simulations is about $1.4\times10^3\,M_\odot$ per gas cell, corresponding to the highest resolution runs in \citet{2019MNRAS.489.4233M}. 
The gravitational softening for star particles is fixed at 7.2\,pc, while for gas cells it is adaptively set with a minimum softening length of 3.6\,pc.
By varying the local star formation efficiency ($\epsilon_{\rm{ff}}$) and enabling/disabling different stellar feedback channels, six distinct subgrid models were simulated, as listed in Table~\ref{tab-1}.

In the following section, we present the properties of coronal gas in our simulated galaxies. We emphasize that no coronal component was included in the initial conditions; all coronal gas originates from internal galactic processes and is subsequently ejected. 
Throughout our analysis, we make no distinction between the ISM and CGM, defining hot halo gas as all gas cells at $T>10^6$\,K within the virial radius ($R_{\rm vir} = 260$\,kpc). This hot component constitutes a minor fraction ($\lesssim10\%$) of the total gas mass within $R_{\rm vir} $.

\begin{deluxetable}{cccc}
\setlength{\tabcolsep}{12pt} 
\renewcommand{\arraystretch}{1.1} 
\tablecaption{Summary of the six subgrid model variations analyzed in this work. \label{tab-1}}
\tablewidth{0pt}
\tablehead{
\colhead{Run} & \colhead{$\epsilon_{\rm ff}$} &  \colhead{Radiation \& Winds} & \colhead{SN} 
}
\startdata
 SFE1 	& 0.01 	& \ding{51} 	& \ding{51}   \\
 SFE10 	& 0.1 	& \ding{51} 	& \ding{51}   \\
 SFE100 	& 1.0 	& \ding{51} 	& \ding{51}   \\
 Nofeed 	& 0.01 	& \ding{55}	& \ding{55}   \\
 Rad 	& 0.01 	& \ding{51} 	&  \ding{55}  \\  
 SN 		& 0.01 	& \ding{55} 	& \ding{51}   \\
 \enddata
 \tablecomments{Columns: (1) simulation run name; (2) star-formation efficiency; (3) inclusion of radiative feedback and stellar winds;  (4) inclusion of SN feedback. SFE1 is the fiducial model with all stellar feedback channels enabled \citet{2019MNRAS.489.4233M}.}
\end{deluxetable}

\section{Results} 
\label{sec:result}

The diffuse X-ray emission depends on the gas density and temperature and is characterized by its emissivity. For hydrogen-dominated gas, the emissivity $\epsilon_\nu$ (in units of $\rm{erg\ s^{-1}\ cm^{-3}\ Hz^{-1}}$) is given by
\be
\epsilon_\nu =  n_e n_{\rm H} \Lambda_\nu(T,\,Z), 
\label{eq-eps}
\ee
where $\nu$ is the frequency of emitted photons, $n_e$ is the electron density, $n_{\rm H}$ is the hydrogen density, and $\Lambda_\nu$ is the cooling function for gas of temperature $T$ and metallicity $Z$. 
The cooling function is dominated primarily by metal-line emission, with a secondary contribution from free--free emission (bremsstrahlung).  
The emissivity table is created using the Astrophysical Plasma Emission Code ({\sc Apec}) model \citep{2001ApJ...556L..91S} within {\sc AtomDB} 3.0.9\footnote{http://www.atomdb.org/}, assuming an optically thin plasma in collisional ionization equilibrium.  
Assuming solar metallicity \citep[$Z=Z_\odot$; e.g.,][]{2013ApJ...773...92H, 2018ApJ...862...34N}, the cooling function in the $0.5-2$\,keV band peaks between $10^6$ and $10^7$\,K, and declines at both higher and lower temperatures\footnote{Setting $Z=0.3Z_\odot$ reduces 0.5--2\,keV luminosity and surface brightness by approximately a factor of 3, and increases the emission measure by nearly the same factor.} . 
The band-integrated emissivity is then given by
\be
\epsilon_X = \int \epsilon_\nu d\nu.
\ee

\begin{figure*}[htbp]
\centering
\includegraphics[width=1\textwidth]{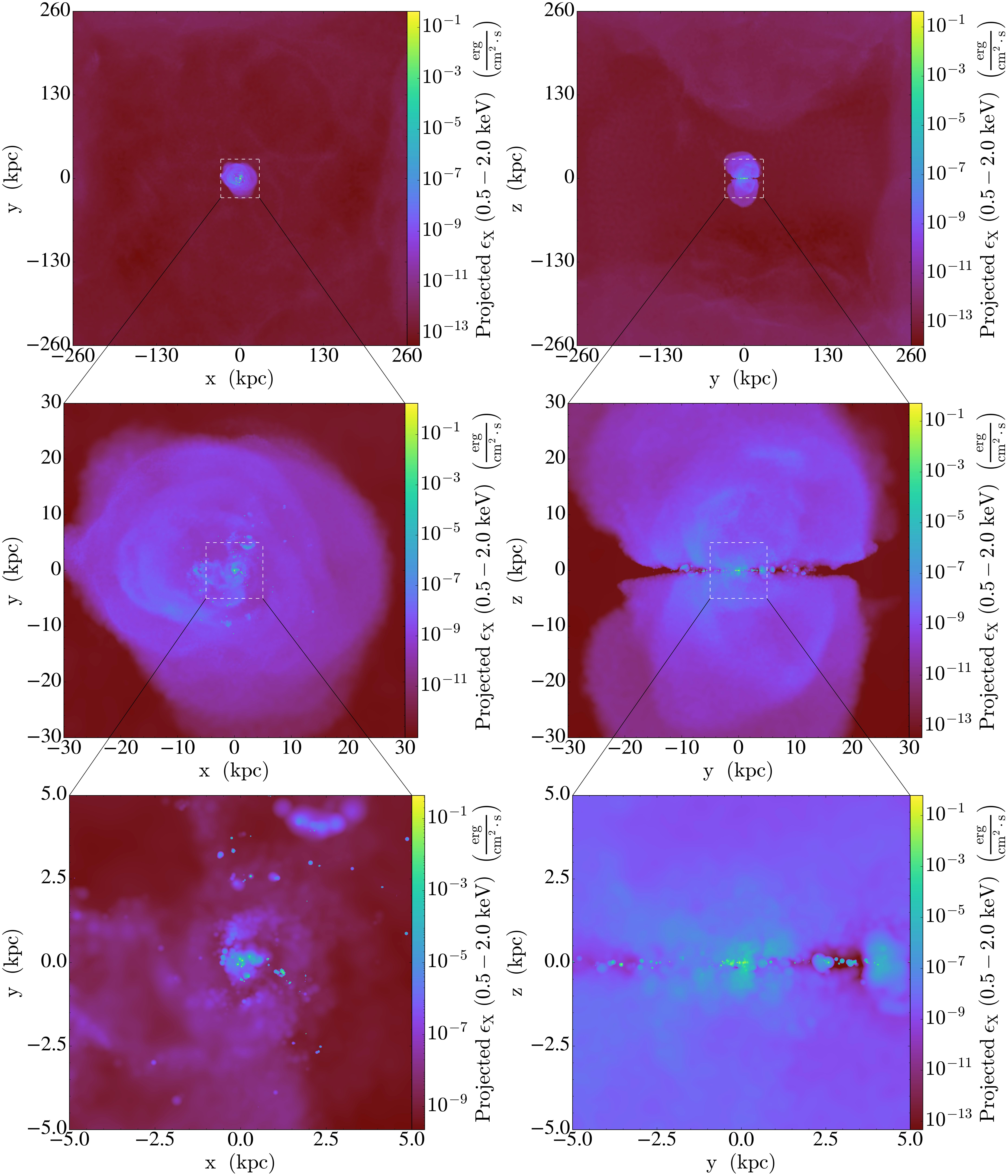}
\caption{Projected $0.5-2$\,keV band emissivity along the $z$-axis (face-on; {\it left panels}) and the $x$-axis (edge-on; {\it right panels}) for the simulated galaxy in the SFE1 run at 0.775\,Gyr.  Rows show progressively zoom-in views (indicated by the white squares in the preceding row), with field-of-view side length of 520\,kpc (top row), 60\,kpc (middle row), and 10\,kpc (bottom row), respectively.
\label{fig-emis}}
\end{figure*}

Figure~\ref{fig-emis} shows face-on and edge-on projections of the $0.5-2$\,keV emissivity for the simulated galaxy in the final snapshot ($t=0.775$\,Gyr) of the SFE1 run, with zoom-ins on the central regions. As shown, the projected emissivity spans more than 10 orders of magnitude, with the  highest values on average occurring in the central region. 
While diffuse features trace superbubbles driven by SNe, the embedded pointlike sources primarily arise from early feedback (stellar winds and/or radiation) from massive stars.  
In the edge-on view, the projected emissivity is low in the disk plane ($z\sim0$) for $y\gtrsim 10$\,kpc  due to the scarcity of hot gas. This results from the declining stellar density at larger radii, which reduces feedback from massive stars and SNe, leaving the gaseous disk predominantly cold.  
Meanwhile, stellar feedback tends to channel energy and momentum more efficiently along the vertical direction than radially.
The mushroom-shaped morphology indicates that SN feedback can extend to $\sim30$\,kpc above and below the galactic plane. 
 
To compare with observations of X-ray emission from the MW and external disk galaxies, we derive the properties of hot gas surrounding our simulated galaxies using the {\tt yt} analysis toolkit\footnote{https://yt-project.org} \citep{2011ApJS..192....9T}. 
Unless otherwise stated, the X-ray properties discussed hereafter (e.g., luminosity and surface brightness) refer to the $0.5-2$\,keV band and analyses are performed using the final snapshot of the simulation run.

\subsection{X-ray luminosity}
\label{sec:luminosity}

The X-ray luminosity of a selected region is calculated as\footnote{Luminosity is calculated using all gas particles within the selected region, including those in background grids. We confirmed that this background gas contributes only $\sim$0.01\%--3\% percent of the time-averaged total luminosity, depending on the subgrid model.}
\be
L_{\rm X} = \sum_i \epsilon_X \frac{m_i}{\rho_i},
\ee
where  $\epsilon_X$, $m_i$ and $\rho_i$ are the emissivity, mass, and density of the $i$th gas particle within the region, respectively.
To enable a direct comparison with observations of external galaxies \citep{2013MNRAS.428.2085L}, we calculate the luminosity within a cylindrical region centered on the galactic center. 
This cylinder extends vertically $\pm 5$\,kpc  from the disk midplane, corresponding to the median extent used for spectral analysis and luminosity calculation in the observed sample.

\begin{figure*}[htbp]
\centering
\includegraphics[width=0.9\textwidth]{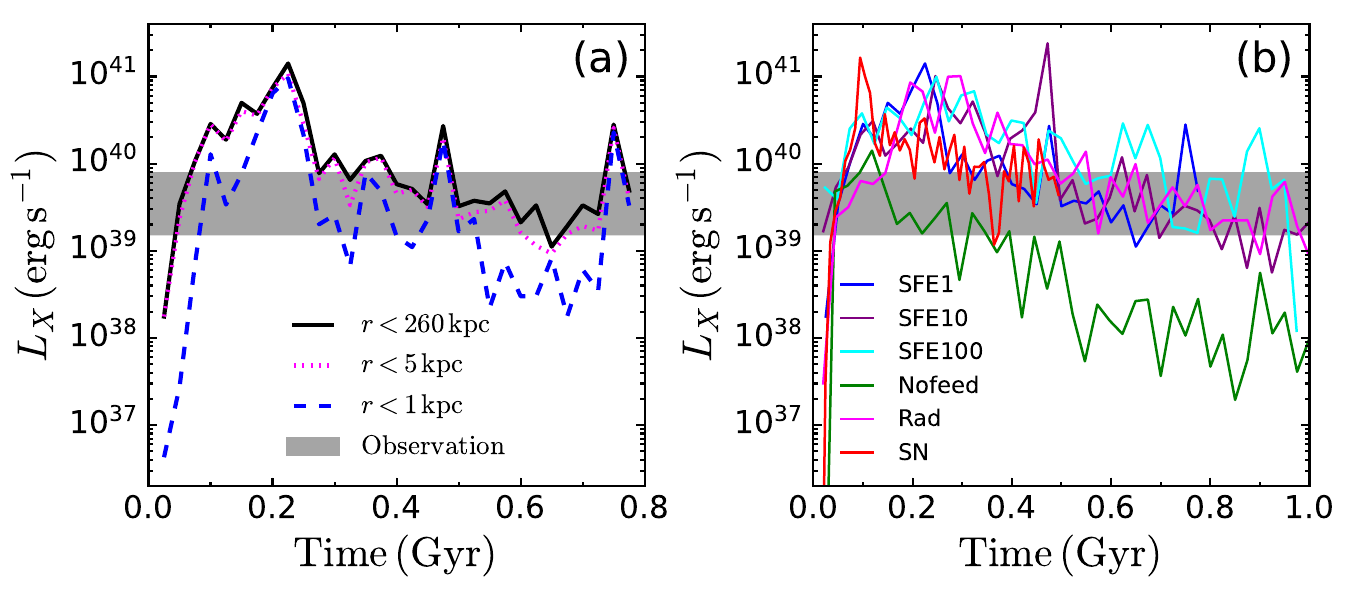}
\caption{{\bf (a)} Evolution of X-ray luminosity ($L_{\rm X}$) within cylindrical regions (height = 10\,kpc) of radii 260, 5, and 1\,kpc for the simulated galaxy in the SFE1 run. The gray band indicates the observed $L_{\rm X}$ range for disk galaxies with stellar mass and SFR comparable to the MW \citep[][]{2012MNRAS.426.1870M, 2016MNRAS.457.1385W}; the MW itself resides within  this band.
{\bf (b)} Time evolution of $L_{\rm X}$ within a cylindrical region (height = 10\,kpc, radius =  260\,kpc) for different simulation runs. 
\label{fig-lx}} 
\end{figure*}

Figure~\ref{fig-lx}(a) shows the evolution of X-ray luminosity within cylindrical regions of different radii for the fiducial SFE1 simulation. The luminosity integrated within the virial radius (260 kpc; black solid line) varies temporally and ultimately converges to the observed value for the MW \citep[$L_{\rm X} = 2.0^{+3.0}_{-1.2} \times10^{39}\,{\rm erg\, s^{-1}}$;][]{1997ApJ...485..125S, 2015ApJ...800...14M} and for external disk galaxies with $M_*$ and SFR comparable to the MW (gray band). 
The luminosity evolution generally traces the star formation history, exhibiting an initial sharp rise following the first episode of gravitational runaway collapse of molecular gas \citep[see Fig. 2 of][]{2020MNRAS.499.5862L}. 
The luminosity within the central 1\,kpc also evolves (blue dashed line) and typically contributes $\sim$10\%--50\% of the total luminosity. 
The instances of the highest fractional contribution (nearly 50\%) predominantly coincide with the sharp peak in the total luminosity.
Since X-ray photons are primarily generated in the central region associated with stellar activity, the central concentration diminishes as activity subsides, allowing the emission to diffuse outward.
Nevertheless, the diffuse X-ray emission remains concentrated within the central $\sim5$\,kpc (magenta dotted line). 
Notably, the total hot gas content is dominated  by material beyond 5\,kpc due to the large volume of these lower-density regions, i.e., hot gas at $5<r<260$\,kpc constitutes about 99.5\% of the total hot gas mass within the virial radius (260\,kpc).

Figure~\ref{fig-lx}(b) shows the evolution of X-ray luminosity within the cylindrical region extending to $r=260$\,kpc for the various simulation runs. All runs exhibit an initial sharp increase in X-ray luminosity, tracking the evolution of the SFR, triggered by the first gravitational runaway collapse of molecular gas occurring after roughly one dynamical time.  Subsequently, $L_{\rm X}$ oscillates with a declining trend. 
The {\it Nofeed} run exhibits significantly lower luminosities on average compared to the other runs, deviating from the observed range (gray band) by 1--2 orders of magnitude. The other runs show no significant differences among themselves. However,  the SFE100 run, which has the highest star formation efficiency, displays violent oscillations, particularly at $\gtrsim0.9$\,Gyr, crossing both above and below the observed range.

\begin{figure*}[htbp]
\centering
\includegraphics[width=0.9\textwidth]{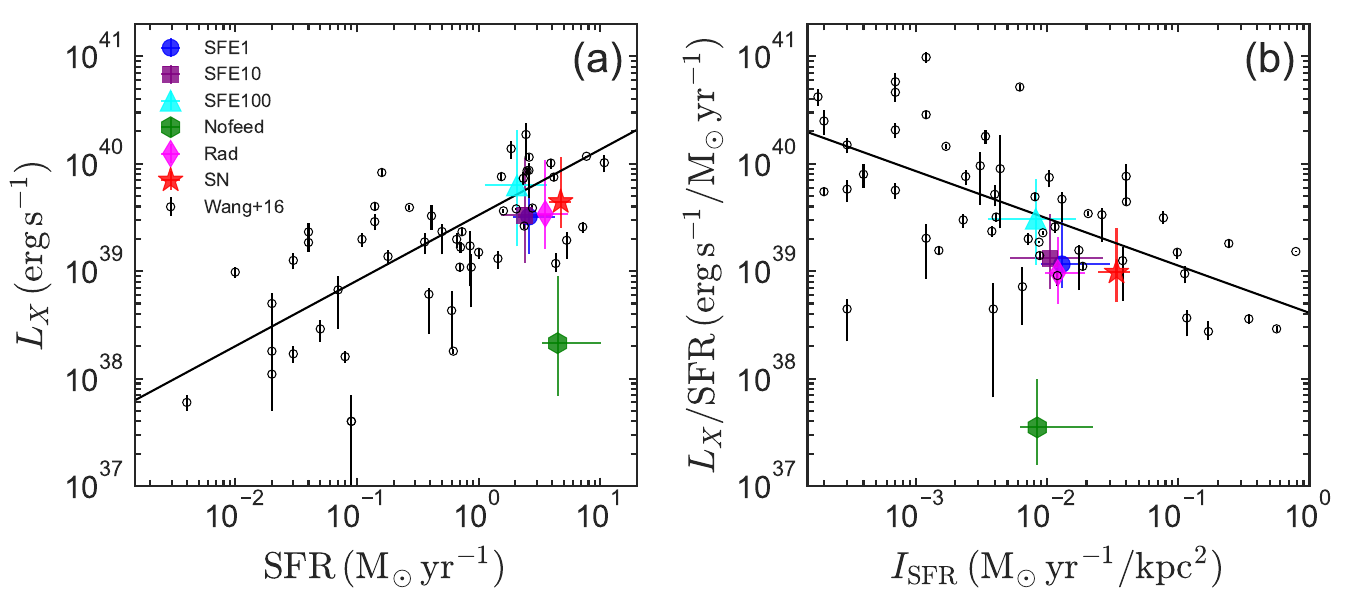}
\caption{Comparison of model predictions (colored symbols) with observed scaling relations. 
{\bf (a)} $L_{\rm X}$--SFR  relation for highly inclined disk galaxies in the local Universe \citep[][]{2016MNRAS.457.1385W}. Black circles show observational data; the solid line is the best-fit log-linear relation. Model $L_{\rm X}$ values were computed within cylinders of virial radius.
The $1\sigma$ uncertainties (derived from the $16$th and $84$th percentiles) use snapshots at $t > 0.4$\,Gyr.
{\bf (b)}: $L_{\rm X}/$SFR versus $I_{\rm SFR}$ relation for the same galaxy sample.
\label{fig-relation}}
\end{figure*} 

We also examine whether our simulated galaxies follow X-ray scaling relations similar to those observed in external disk galaxies. To calculate the medians and uncertainties of X-ray luminosity and SFR, we consider only snapshots at $t>0.4$\,Gyr and adopt a cylindrical radius of $260$\,kpc. The time cutoff ensures the gaseous disk has settled into a new equilibrium configuration and the multiphase ISM is fully developed. The SFR for each snapshot is averaged over 5\,Myr.

Figure 3(a) displays the results from our simulation runs overplotted on the observed
$L_{\rm X}$--SFR relation for nearby highly inclined disk galaxies from \citet{2016MNRAS.457.1385W}.
We find that the median SFR for all six runs falls within $\sim2-5\,M_\odot\,{\rm yr^{-1}}$, consistent within uncertainties with observations of both the MW and external disk galaxies. 
While the {\it Nofeed} run underpredicts the diffuse X-ray luminosity by $\sim1$\,dex, the X-ray luminosities predicted by the other runs generally agree with the observed relation (solid line), despite the large scatter of the relation. 
This agreement suggests that although early feedback mechanisms (stellar winds and radiation) from massive stars primarily heat gas on smaller spatial scales, they can ultimately produce X-ray luminosities comparable to those generated by SN feedback. 

Figure~\ref{fig-relation}(b) compares our simulation results with the observed $L_{\rm X}/$SFR versus $I_{\rm SFR}$ relation for the same galaxy sample. Here $I_{\rm SFR} \equiv {\rm SFR}/(\pi R^2_{90})$ represents the SFR surface density, where $R_{90}$ is the radius  enclosing 90\% of the total SFR within a cylinder of height 20\,kpc centered on the galaxy. 
Again, the {\it Nofeed} run is an outlier, exhibiting $L_{\rm X}/$SFR values $\sim$1--2 orders of magnitude lower than observed. The other runs, however, show general agreement with the observed relation \citep{2016MNRAS.457.1385W}.  Our results suggest that while the subgrid model lacking feedback can be ruled out, the remaining subgrid models cannot be unambiguously distinguished based solely on these X-ray scaling relations.

\subsection{X-ray surface brightness}
\label{sec:sx}

For an arbitrary line of sight, the surface brightness is obtained by integrating the emissivity along the path length $s$:  
\be
S_X = \frac{1}{4\pi}\int \epsilon_X ds, 
\label{eq-sb}
\ee
where the integration extends from 0 to 260\,kpc, and $S_{\rm X}$ has units of ${\rm erg\ s^{-1}\ cm^{-2}\ sr^{-1}}$.
To compare with the MW observations, we place an observer at the Sun's location, which is $8.2$\,kpc from the galactic center in the disk plane \citep{2016ARA&A..54..529B}. 
Specifically, the observer's coordinates are ($x_0+8.2$, $y_0$, $z_0$), where ($x_0$, $y_0$, $z_0$) denote the galactic center coordinates in units of kiloparsecs. The lines of sight  are defined by their Galactic coordinates ($l$, $b$).

\begin{figure*}[htbp]
\centering
\includegraphics[width=0.9\textwidth]{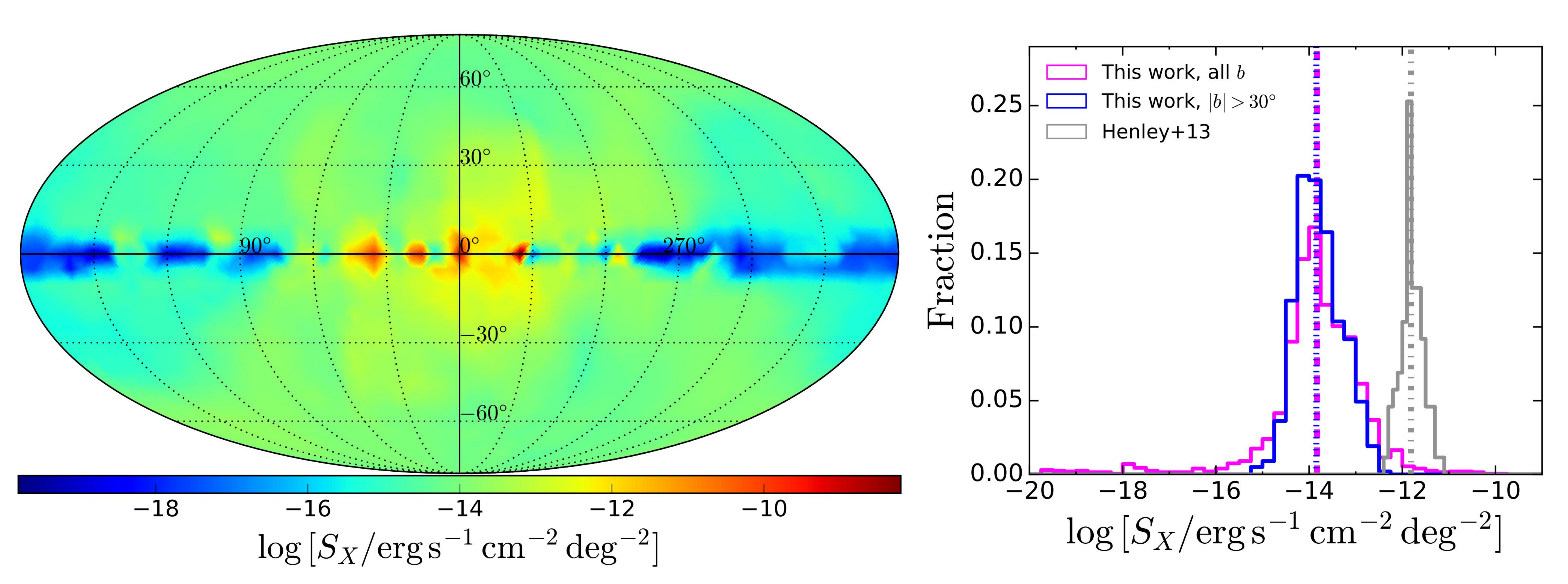}
\caption{{\bf Left:} all-sky X-ray surface brightness in Mollweide projection for an observer at the solar position. 
{\bf Right:} comparison of the surface brightness distributions with XMM-Newton observations of the MW \citep[gray histogram; 110 sightlines with $|b|>30^\circ$;][]{2013ApJ...773...92H}. The magenta and blue histograms show simulated results for all sightlines and $|b|>30^\circ$ sightlines, respectively.  
Lines of corresponding colors indicate the median values.
\label{fig-sb}}
\end{figure*}

The resulting all-sky map of the surface brightness is shown in the left panel of Figure~\ref{fig-sb}. 
This map was generated by first dividing the entire sky into a grid with $\Delta l = 5^\circ$ and $\Delta b = 5^\circ$, calculating $S_X$ for each grid line of sight, and then linearly interpolating to obtain values on a finer grid with $0.5^\circ$ resolution. 
As shown, $S_X$ varies significantly across different lines of sight, differing by $\gtrsim10$ orders of magnitude, consistent with the emissivity distribution shown in  Figure~\ref{fig-emis}. The highest and lowest surface brightness both occur near the disk plane: toward the galactic center ($l\sim0^\circ$) and anticenter ($l\sim180^\circ$) directions, respectively. 
This behavior directly reflects the projected emissivity results in Fig.~\ref{fig-emis}. The X-ray emission originates predominantly from the central few kiloparsecs due to the higher density of hot gas. Consequently, for an observer at the Sun's location, only sightlines passing through these central regions exhibit high surface brightness. 

To mitigate potential biases arising from using a single observer at a specific location, we place four observers at off-center positions. Each observer is located $8.2$\,kpc from the galactic center, with their positions separated by $90^\circ$ azimuthally \citep[e.g.,][]{2020ApJ...896..143Z, 2024ApJ...962...15Z}. 
For each observer, we generate 500 random lines of sight and calculate the corresponding surface brightness. The Galactic longitude $l$ is generated uniformly over  $[0^\circ, 360^\circ)$, while the latitude $b$ is generated over [$-90^\circ, 90^\circ$] with a probability density proportional to $\cos(b)$, i.e., $P(b) \propto \cos(b)$.

The right panel of Figure~\ref{fig-sb} displays the distribution of surface brightness for a total of 2000 sightlines (500 per observer $\times$ 4 observers). Unlike the narrow distribution ($\sim1$\,dex) observed for the Galactic halo \citep[gray histogram;][]{2013ApJ...773...92H}, the simulated galaxy's surface brightness exhibits a strong dependence on the sightline orientation (magenta histogram). 
Sightlines penetrating the central region (e.g., $|b|\lesssim10^\circ$ and $|l|\lesssim30^\circ$\footnote{In our notation, this longitude interval corresponds to $l \lesssim 30^\circ$ or $l \gtrsim 330^\circ$.}) show brightness comparable to or even exceeding the observed ranges.
In contrast, antigalactic sightlines can exhibit surface brightness 4--8 orders of magnitude lower than the observed values.  
When applying observational constraints such as $|b|>30^\circ$ (blue histogram), both the brightest and darkest sightlines are excluded. This results in a median value nearly identical to the unconstrained case; however, our median results under these constraints remain $\sim2$ orders of magnitude lower than the observations.

\begin{figure*}[htbp]
\centering
\includegraphics[width=0.98\textwidth]{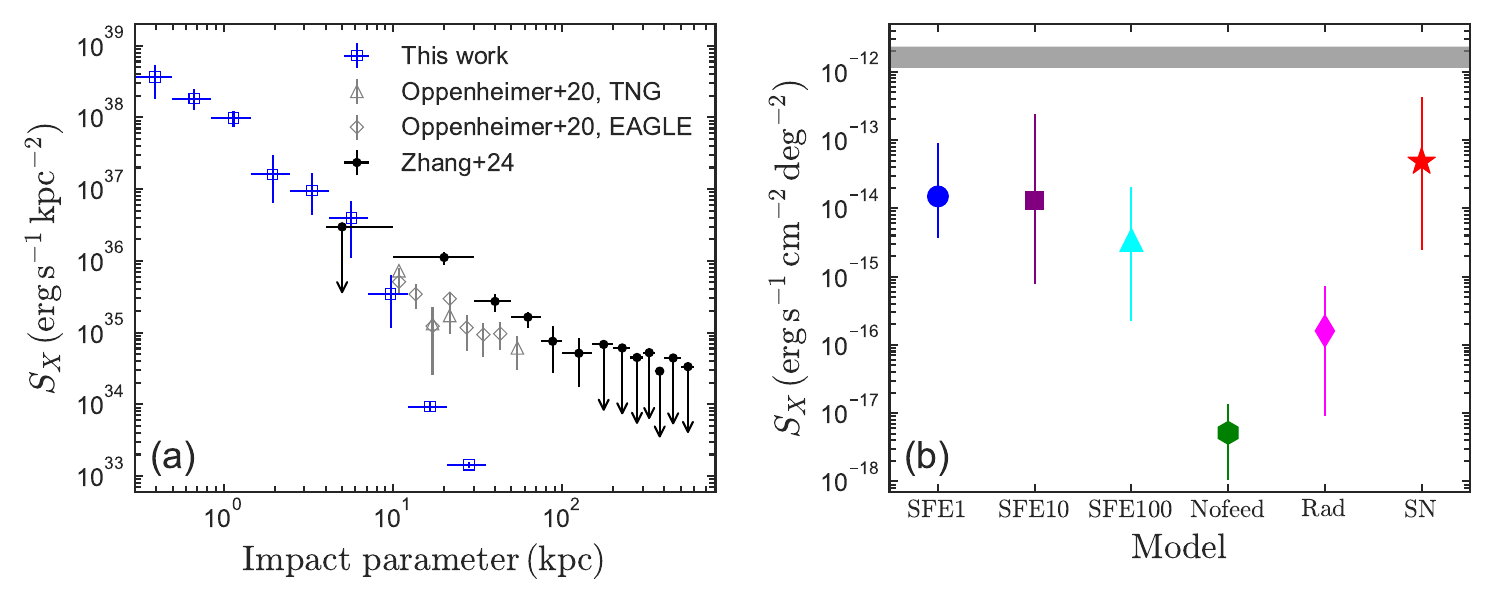}
\caption{{\bf (a)} Surface brightness profile for the SFE1 run compared to eROSITA observations of MW-mass stacks \citep[black circles;][]{2024A&A...690A.267Z}. 
Results from EAGLE (gray diamonds) and IllustrisTNG (gray triangles) for low-mass, low-sSFR galaxies are overplotted \citep[see][for sample selection criteria]{2020ApJ...893L..24O}. 
{\bf (b)} Median surface brightness with $1\sigma$ uncertainties for different subgrid models (solar perspective). The gray band shows the observed range for the Galactic halo \citep{2013ApJ...773...92H}.
\label{fig-sx5}}
\end{figure*} 

To compare with the stacking results for approximately 80,000 MW-mass galaxies from eROSITA \citep{2024A&A...690A.267Z}, we present the surface brightness profile from external views in Figure~\ref{fig-sx5}(a). This profile was generated similarly to the observational methodology but incorporates various galaxy orientations. 
In our simulations, the surface brightness generally declines with increasing radius. The profile steepens beyond $r\gtrsim5$\,kpc compared to the inner region.  
This results in an underestimation of the surface brightness in the outer regions ($r\gtrsim10$\,kpc) by at least 2 orders of magnitude, consistent with the inside-out view result shown in Fig.~\ref{fig-sb}(b). 
As observational data for the inner regions are  unavailable, a direct comparison in this radial range is not possible.

Figure~\ref{fig-sx5}(b) compares the surface brightness for various subgrid models, from a solar perspective (similar to Fig.~\ref{fig-sb}). Simulations including SN feedback align more closely with observational determinations, largely owing to the large-scale superbubbles driven by SN explosions. However, even for the best-matching SN run, a discrepancy of $\sim1.5$~dex remains when comparing median values.

 \subsection{Emission Measure and Temperature}  
 \label{sec:em}
 
The emission measure (EM), along with the temperature ($T$), determines the emissivity. EM and $T$ can be simultaneously derived by fitting X-ray spectra with plasma models (e.g., {\sc Apec}), often under the simplifying  assumption of a single-temperature ($1T$) component, despite significant temperature variations potentially existing along the sightline.  

The emission measure is calculated as
\be
{\rm EM}=\int n_e n_{\rm H} ds,
\label{eq-em}
\ee
where the integration extends from 0 to 260\,kpc.
 
To make a relatively straightforward comparison with observations, we adopt the emissivity-weighted temperature along the sightline as the ``measured" temperature  \citep[e.g.,][]{2020ApJ...904L..14G}. 
Given that $1T$ models typically yield $T\gtrsim10^6$\,K for the Galactic hot halo, we calculated EM by considering only gas particles with $T>10^{6}$\,K \citep[e.g.,][]{2015ApJ...800..102H}.

\begin{figure*}[htbp]
\centering
\includegraphics[width=0.9\textwidth]{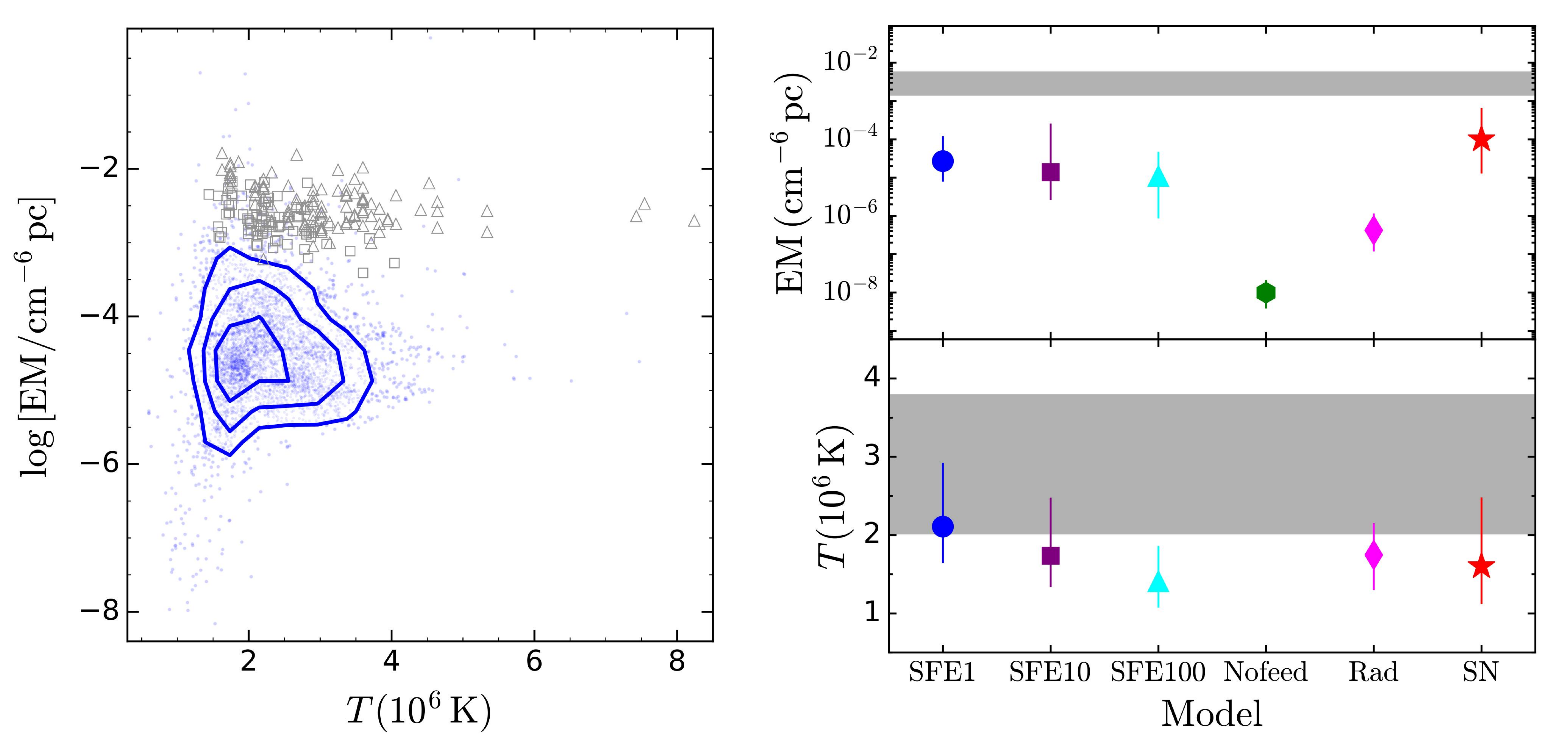}
\caption{{\bf Left:} emission measure versus emissivity-weighted temperature.  Blue dots show results from the SFE1 run; blue contours indicate the $1\sigma$--$3\sigma$ confidence levels. Gray symbols represent observations: squares from XMM-Newton \citep{2013ApJ...773...92H} and triangles from Suzaku \citep{2018ApJ...862...34N}. 
{\bf Right:} predicted emission measure and temperature for different simulation runs. The gray band marks the observed $1\sigma$ range from combined XMM-Newton and Suzaku data. 
\label{fig-emT}}
\end{figure*}

Figure~\ref{fig-emT}(a) displays the EM--$T$ distribution for 5000 random sightlines originating from four off-center observers in the SFE1 run. 
The predicted temperature distribution (blue dots and contours) is broadly consistent with  observations \citep[gray squares and triangles;][]{2013ApJ...773...92H, 2018ApJ...862...34N}, with median temperatures of $2.1^{+0.8}_{-0.5} \times10^6$\,K  and $3.0^{+0.8}_{-0.1} \times10^6$\,K, respectively.
However, the EM values are systematically underestimated by $\sim2$ orders of magnitude. 
The discrepancy arises because denser hot gas particles are predominantly concentrated within the central $\sim5$\,kpc, causing most random sightlines from off-center observers to avoid this high-density region.  

The EM--$T$ distributions for other simulation runs are similarly derived and compared in Figure~\ref{fig-emT}(b). The dependence of EM on subgrid models follows a trend analogous to that of surface brightness in Fig.~\ref{fig-sx5}(b), as expected since both quantities scale with the square of gas density.  Furthermore, all simulated EMs fall below observational constraints \citep[gray band;][]{2013ApJ...773...92H, 2018ApJ...862...34N}. 
Temperatures from most models (SFE1, SFE10, Rad, and SN) remain consistent with observations within uncertainties. The {\it Nofeed} run is excluded due to insufficient hot gas production, rendering its inferred temperature physically unmeaningful.

\section{Discussion} 
\label{sec:discuss}

\subsection{Comparison with the literature} 

 In Section~\ref{sec:result}, we  analyzed high-resolution simulations of isolated galaxies from both internal (inside-out) and external perspectives. The results demonstrate that the SMUGGLE model broadly reproduces the X-ray luminosity of MW-mass disk galaxies across various stellar-feedback subgrid models. However, X-ray emission predominantly originates from the central $\sim5$\,kpc region. This spatial concentration leads to underestimations of surface brightness and emission measure by $\sim2$ orders of magnitude for inside-out sightlines. 

Similar discrepancies were reported by \citet{2015ApJ...800..102H}, whose magnetohydrodynamical model of SN-driven ISM \citep{2012ApJ...750..104H} yielded comparable results.
They proposed that key physical processes such as charge exchange emission, cosmic-ray-driven outflows, and accretion-powered extended gaseous halos may be essential for reconciling simulations with observations.

Additional studies have compared cosmological simulations with X-ray observations of MW-mass systems. These typically employ mock X-ray observations of individual galaxies or galaxy stacks from external views \citep[e.g.,][]{2020MNRAS.494..549T, 2021MNRAS.502.2934K}. 
For instance, \citet{2020ApJ...893L..24O} detected diffuse X-ray emission in the CGM of low-sSFR MW-mass galaxies within the EAGLE \citep{2015MNRAS.446..521S} and IllustrisTNG \citep{2018MNRAS.473.4077P} simulations. The derived X-ray luminosity within $10-200$\,kpc reaches $L_{\rm X}\sim 10^{39}\,{\rm erg\,s^{-1}}$, approximately half the value measured for the MW \citep[][]{1997ApJ...485..125S}.  
Furthermore, both simulations underpredict  the eROSITA stacked surface brightness of MW-mass galaxies \citep[][]{2024A&A...690A.267Z} by factors of $\sim5-10$ (gray diamonds and triangles in Fig.~\ref{fig-sx5}(a)).

\citet{2020MNRAS.494..549T} conducted mock X-ray observations of galaxies across a broad mass range in the IllustrisTNG simulations. They found that the $0.3-5$\,keV luminosity within $R_e$ and $5R_e$ for MW-mass galaxies is consistent with Chandra observations of late-type galaxies \citep{2012MNRAS.426.1870M, 2013MNRAS.428.2085L}. 
Additionally, they confirmed the dependence of X-ray luminosity on SFR in low-mass galaxies and reported a systematic difference between star-forming and quenched systems: the former exhibits X-ray luminosities approximately 1 order of magnitude  higher on average. They attributed this disparity to galaxy quenching driven by  kinetic AGN feedback from supermassive black holes -- a process absent in our current simulations.  

\citet{2021MNRAS.502.2934K} investigated the origin of X-ray coronae in central disk galaxies using the EAGLE simulations \citep{2015MNRAS.446..521S}. Their results show good agreement between simulated X-ray luminosities of MW-mass galaxies and  observational data. 
Furthermore, they demonstrated that while SN-heated gas dominates the X-ray emission within the central region ($\le 0.1R_{\rm vir}$), a quasi-hydrostatic, accreted atmosphere prevails beyond this radius for galaxies of mass $\le 10^{12}\,M_\odot$.

Including a galactic hot corona may resolve the tension between our results and observations. 
The hot corona plays a fundamental role in galaxy evolution. When implemented in  initial conditions  analogous to our simulations \citep[e.g.,][]{2023MNRAS.524.4091B, 2025A&A...697A.121B}, such coronae establish a gas reservoir essential for sustaining star formation. This process drives a gas circulation cycle (the galactic fountain) between the disk and corona, mediated by stellar feedback. We reserve analysis of corona-embedded simulations for future investigation.

\subsection{The Impact of Cosmic Rays} 

While gas heating via cosmic-ray interactions is implemented in the SMUGGLE model, this effect primarily applies for dense gas. 
Recently, \citet{2025arXiv250118696H} proposed that inverse Compton scattering of cosmic microwave background (CMB) photons by GeV cosmic-ray electrons could be responsible for the diffuse X-ray emission in low-mass systems. 
Here, we examine whether incorporating this effect can improve the agreement of the SMUGGLE model with observations. 

\begin{figure*}[htbp]
\centering
\includegraphics[width=0.9\textwidth]{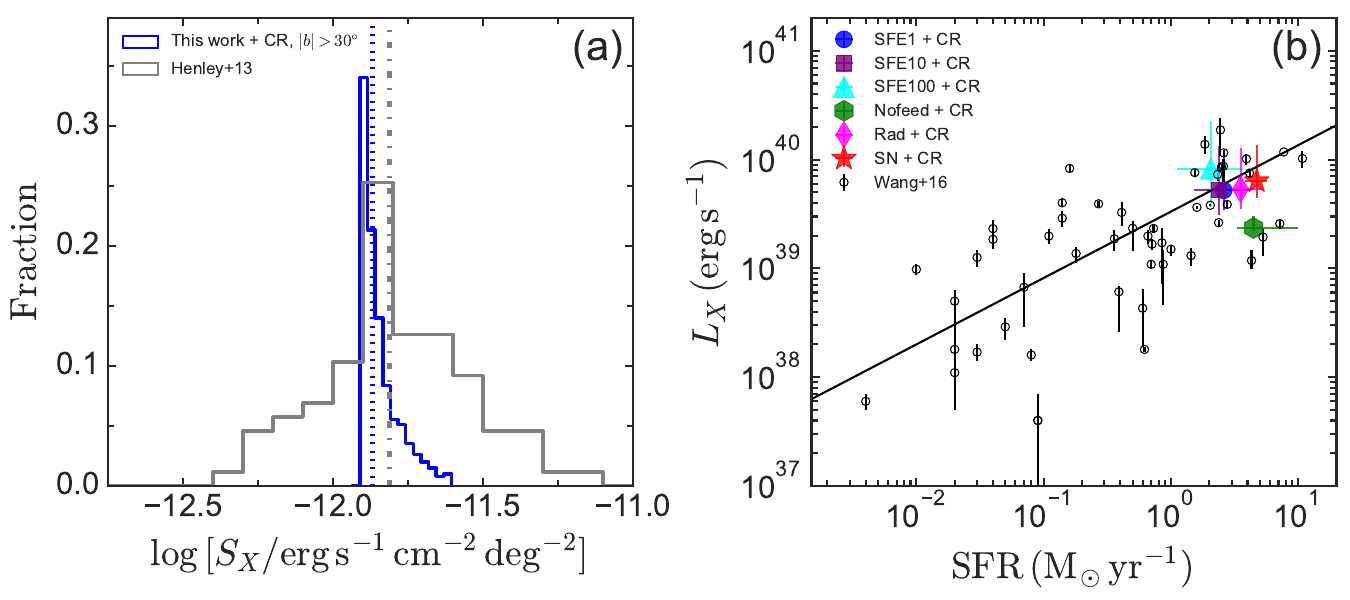}
\caption{{\bf (a)} Surface brightness distribution (methodology identical to Figure~\ref{fig-sb}(b)) including contributions from inverse Compton scattering of CMB photons by cosmic rays (blue). 
{\bf (b)} $L_{\rm X}$--SFR relation (cf. Figure~\ref{fig-relation}) incorporating cosmic ray effects.
\label{fig-sb_cx}}
\end{figure*}

Figure~\ref{fig-sb_cx}(a) shows surface brightness distribution for the SFE1 run, now including cosmic-ray-induced hot gas via the inverse Compton process (blue histogram). 
We adopt a leptonic energy injection rate of $\dot{E}_{40} \equiv \dot{E}_{{\rm cr, } l}/10^{40}\,{\rm erg\, s^{-1}} = 0.2$ and a streaming speed of $v_{100} \equiv v_{\rm st, eff}/100\,{\rm km\,s^{-1}} = 1.0$ \citep[see Eq.(6) of][]{2025arXiv250118696H}. The results from their Figures 2 and Fig. 4 are then rescaled to update the surface brightness and luminosity according to these parameters.
While the median surface brightness now shows good agreement with observations (vertical lines), our results exhibit a narrower distribution dominated by cosmic-ray effects. 

Figure 7(b) presents the $L_{\rm X}$--SFR relation, now incorporating cosmic-ray effects. The results remain consistent with Chandra observations of nearby disk galaxies and, in fact, show improved agreement compared to those presented in Fig.~\ref{fig-relation}(a).

However, the agreement of both the surface brightness distribution and the total luminosity with observations does not necessarily imply that cosmic rays are the sole or  correct solution. Instead, this demonstrates that any mechanism capable of contributing to the development of a hot CGM, such as AGN feedback and/or gas accretion from the IGM, may help reconcile the model with the observations. 
Crucially, while the surface brightness increases by $\sim2$\,dex in this scenario, the total X-ray luminosity remains consistent with the observed values.

\section{Summary} 
\label{summary}

We study diffuse X-ray emission from hot gaseous halos in MW-like galaxies by analyzing a series of high-resolution simulations using the SMUGGLE galaxy formation model. Our analysis incorporates both inside-out (solar location) and external perspectives. The main conclusions are as follows.

\begin{enumerate}[label=(\roman*), align=left]
	\item For the five subgrid models with stellar feedback included, the overall X-ray luminosities agree with MW observations and scaling relations for external disk galaxies. This suggests that early feedback (stellar wind and radiation) from massive stars contributes X-ray luminosities comparable to those from SN explosions, despite its more localized impact.
	\item Viewed from the solar location, all simulations underpredict the median surface brightness of the MW by $\sim1-5$ orders of magnitude.
This occurs because the simulated denser hot gas is concentrated within the central $\sim 5$\,kpc, leaving most random sightlines devoid of luminous X-ray-emitting regions. This  radial concentration is confirmed by the external view: the simulated surface brightness profile underestimates observations of  MW-mass galaxy stacks  beyond $\sim10$\,kpc by $\gtrsim2$ orders of magnitude. 
	\item The emissivity-weighted temperatures across simulations generally match MW observations. However, all runs underestimate the EM by $\sim$1--5 orders of magnitude.
	\item Including inverse Compton scattering of CMB photons by cosmic rays  alleviates the tension: the median surface brightness becomes consistent with MW observations while maintaining consistency with observed luminosity.
Incorporating a hot coronal component in the initial conditions may also help. Alternatively, solutions could involve AGN feedback and/or gas accretion from the IGM.
\end{enumerate}

The SMUGGLE model successfully reproduces the kinematics and spatial distributions of warm gas traced by \ion{O}{6} absorption \citep{2024ApJ...962...15Z, 2017ApJ...835...52F, 2020ApJ...893...82F}. The inconsistency between the simulated X-ray coronae and observations indicates that the origins of hot gas (at least partially) differ from those of cooler gas.  Studying absorptions of highly ionized metal species (e.g., \ion{O}{7} and \ion{O}{8}) would help elucidate the detailed effects of different stellar feedback  channels on hot gas. 
Future versions of the SMUGGLE model will incorporate cosmological simulations and serve as a powerful tool for predicting multiphase gas properties.

\begin{acknowledgements}
X.Z. thanks Zhijie Qu for helpful discussions on various aspects of this work. 
This work is supported by the National Natural Science Foundation of China (grant Nos. 11890692, 12133008, 12221003, 12273031, 12192220, 12192223, and 12361161601), the China Manned Space Program (grant Nos. CMS-CSST-2025-A10, CMS-CSST-2021-B02)
the Fundamental Research Fund for the Central Universities of China (grant No. 20720230016),  the Fujian Provincial Natural Science Foundation of China (Grant No. 2024J08001), and the Natural Science Foundation of Xiamen, China (No. 3502Z202472007). 
F.M. acknowledges funding by the European Union--NextGenerationEU, in the framework of the HPC project--“National Centre for HPC, Big Data and Quantum Computing” (PNRR--M4C2--I1.4--CN00000013--CUP J33C22001170001).
QY was supported by the European Research Council (ERC) under grant agreement No. 101040751.
\end{acknowledgements}

\software{ 
yt \citep{2011ApJS..192....9T}, 
Astropy \citep{2018AJ....156..123A}, 
Matplotlib \citep{2007CSE.....9...90H}, 
SciPy \citep{2020NatMe..17..261V}.
}

\bibliographystyle{aasjournal}
\bibliography{ms}{}
\end{CJK*}
\end{document}